\documentclass[twocolumn,noshowpacs,preprintnumbers,amsmath,amssymb]{revtex4-2}
\usepackage{graphicx}
\usepackage{dcolumn}
\usepackage{bm}

\begin{document}

\title{Superconducting THz sources with 12$\%$ power efficiency}
\author{R. Cattaneo$^1$}
\author{E. A. Borodianskyi$^1$}
\author{A. A. Kalenyuk$^{2,1}$}
\author{V. M. Krasnov$^{1,3}$}
\email{Vladimir.Krasnov@fysik.su.se}

\affiliation{$^1$ Department of Physics, Stockholm University,
AlbaNova University Center, SE-10691 Stockholm, Sweden;}

\affiliation{$^2$ Institute of Metal Physics of National Academy
of Sciences of Ukraine, 03142 Kyiv, Ukraine;}

\affiliation{$^3$ Moscow Institute of Physics and Technology,
State University, 9 Institutsiy per., 141700 Dolgoprudny, Russia;}


\begin{abstract}
Low power efficiency is one of the main problems of THz sources,
colloquially known as ``the THz gap". In this work we present
prototypes of THz devices based on whisker-crystals of a
high-temperature superconductor Bi$_2$Sr$_2$CaCu$_2$O$_{8+\delta}$
with a record high radiation power efficiency of $12\%$ at a
frequency of $\sim 4$ THz. We employ various on- and off-chip
detection techniques and, in particular, use the radiative cooling
phenomenon for accurate evaluation of the emission power.
We argue that such devices can be used for creation of tunable,
monochromatic, continuous-wave, compact and power-efficient THz
sources.
\end{abstract}

\maketitle


Tunable, monochromatic, continuous-wave (CW), compact and
power-efficient sources of terahertz (THz) electromagnetic waves
(EMW) are required for a broad variety of applications such as
spectroscopy, environmental control, security, non-ionizing
medical imaging, ultra-high-speed telecommunication and
electronics, as well as for fundamental research in various areas
of science \cite{Tonouchi_2007,Note1}. The key problem of THz
sources, colloquially known as the ``THz gap", is a rapid decay of
radiation power efficiency (RPE), i.e. the ratio of emitted and
dissipated power \cite{Note2}, in the low THz range
\cite{Tonouchi_2007}. Despite a significant progress, achieved in
development of semiconducting quantum cascade lasers (QCL's)
\cite{Slivken_2015,Capasso_2015,Khalatpour_2021}, the RPE of CW
QCL's drops from $\simeq 28~\%$ at $f\simeq 55$ THz
\cite{Lyakh_2016} to a sub-percent at 3-4 THz
\cite{Wang_2016,Curwen_2019} and to $\sim 0.01 \%$ at $f\simeq
1.3$ THz \cite{Walther_2007}. Although QCL frequency can be tuned
\cite{Slivken_2015,Capasso_2015,Curwen_2019,Belkin_2013,Rosch_2015},
this comes at the expense of a dramatic reduction of RPE to $\sim
0.0001~\%$ \cite{Belkin_2013} when low THz emission is obtained by
mixing or downconversion of higher frequencies
\cite{Belkin_2013,Rosch_2015}. QCL's emitting at primary low THz
frequencies, on the other hand, have to be cooled down to
cryogenic temperatures, $k_B T \lesssim hf$
\cite{Wang_2016,Curwen_2019,Walther_2007}.

Superconducting devices, based on arrays of Josephson junctions
(JJ's) have an inherent frequency tunability and provide and
alternative technology for creation of cryogenic THz sources with
tunable, monochromatic CW operation
\cite{Han_1994,Barbara_1999,Koshelets_2000,Ozyuzer_2007,Benseman_2013,Welp_2013,Kashiwagi_2015,HBWang_2015,Borodianskyi_2017,Galin_2018,Sun_2018,Kashiwagi_2018,HBWang_2019,Kuwano_2020,Tsujimoto_2020,Saiwai_2020,Delfanazari_2020,Galin_2020}.
The Josephson frequency,
$f_J=(2e/h)V$,
is proportional to the dc-bias voltage $V$
and is limited only by the superconducting energy gap, which can
be in excess of $30$ THz for high-$T_{\textrm{c}}$ superconductors
\cite{Krasnov_2000,SecondOrder}. Emission with a sub-mW power at
$f\simeq 0.5$ THz was achieved from large-area mesa structures
etched on top of single crystals of a layered cuprate
superconductor Bi$_2$Sr$_2$CaCu$_2$O$_{8+\delta}$ (Bi-2212)
\cite{Benseman_2013}. Such mesas represent stacks of atomic scale
intrinsic JJ's \cite{Kleiner}. Tunable EMW emission at the primary
Josephson frequency in the whole THz range 1-11 THz has been
reported from small-area Bi-2212 mesas \cite{Borodianskyi_2017},
albeit with a lower power. The RPE for both large
\cite{Benseman_2013} and small \cite{Borodianskyi_2017} mesas on
Bi-2212 single crystals is $\lesssim 1\%$, which is decent for THz
sources, but small compared to the theoretical limit of $50\%$
\cite{Krasnov_2010}.
The suboptimal operation is caused by impedance mismatching
between a device and an open space. Another key limitation for all
cryogenic devices is set by self-heating. Taking into account
limited cooling power of compact cryo-refrigerators (sub-Watt at
low $T$), devices with RPE $\sim 1\%$ would not be able to emit
much more than 1 mW. Therefore, further enhancement of the
emission power from compact cryogenic devices may only be achieved
via enhancement of RPE.

\begin{figure*}[t]
    \centering
    \includegraphics[width=0.99\textwidth]{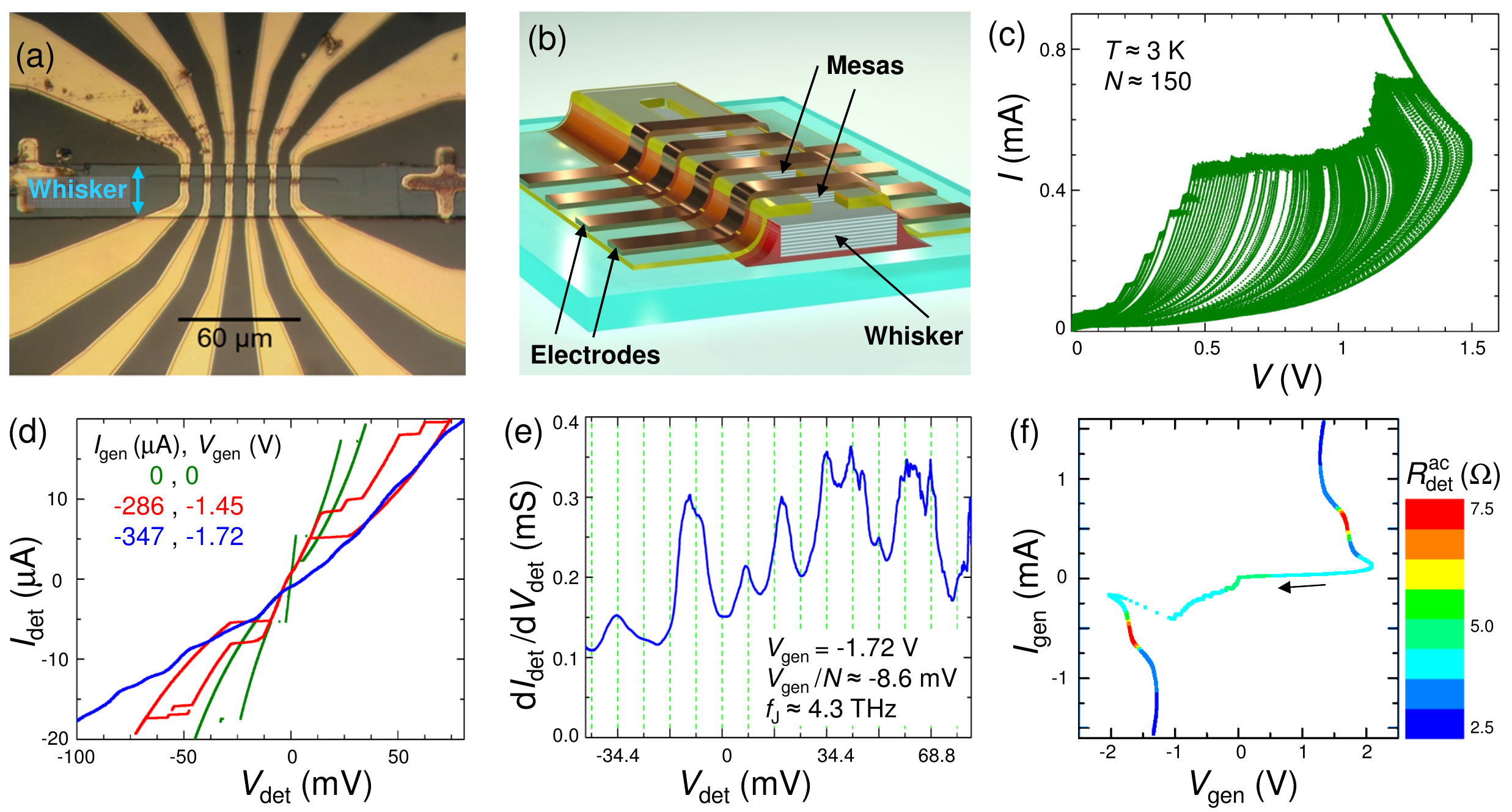}
    \caption{(Color online). (a) Optical image of one of the studied devices.
    Six mesas are formed beneath gold electrodes. (b) A sketch of the device.
    (c) The $I$-$V$ characteristics (integrated oscillogram) of a mesa.
    Multiple branches correspond to switching of individual, $N\simeq 150$, JJ's from
    superconducting to resistive state. (d) $I$-$V$ curves of a detector mesa 
    at three bias points in the generator, marked by circles in Fig. \ref{fig:fig2}
    (a). (e) Differential conductance of the blue $I$-$V$ from (d).
    Peaks represent current steps. Grid lines corresponds to expected Shapiro step
    voltages. (f) On-chip generation-detection experiment with electrically disconnected mesas:
    Here we plot $I$-$V$ of the generator mesa, color mapped by the ac-resistance of electrically separated detector mesa on the same chip. The black arrow indicates the bias sweep direction.
    It is seen that the emission (red color) occurs at steps in the $I$-$V$.}
    \label{fig:fig1}
\end{figure*}

Here we present prototypes of novel THz sources based on Bi-2212
whisker-type crystals with intermediate-size mesa structures. We
employ various techniques for detection of THz radiation such as
an {\em in-situ} detection by a mesa on the same whisker, an
on-chip detection by electrically isolated mesa and an off-chip
detection by a bolometer. Furthermore, we employ the radiative
cooling phenomenon for estimation of the absolute value of the
emitted power. It reveals that the RPE of our devices can reach
$12\%$ making a significant step forward towards the theoretical
limit of $50\%$. The boost of efficiency is attributed to a good
impedances matching with open-space, caused by a specific
turnstile antenna-like geometry of our devices.
We argue that such devices can be used for creation of tunable,
and, most importantly power-efficient THz sources.


Figures \ref{fig:fig1} (a) and (b) show an image and a sketch of studied
devices. In the middle of each device there is a Bi-2212 whisker
crystal with typical sizes $(300-500)\times (20-30) \times (1-5)~
\mu$m$^3$ along crystallographic $a$, $b$ and $c$-axes,
respectively. Several metallic electrodes with the width
$10-15~\mu$m are made across the whisker. Beneath each electrode
there is a mesa structure containing $N\sim 150-250$ intrinsic
JJ's. Devices are made by conventional microfabrication
techniques. Details about sample fabrication and the experimental
setup can be found in Ref. \cite{SecondOrder,Borodianskyi_2017}
and the Supplementary \cite{Supplem}.

Fig. \ref{fig:fig1} (c) shows the current-voltage ($I$-$V$)
characteristics of a mesa at $T\simeq 3$ K. It consists of
multiple branches, due to one-by-one switching of JJ's from the
superconducting to the resistive state
\cite{Kleiner,SecondOrder,Krasnov_2000}. The total number of JJ's,
$N\simeq 150\pm 10$ for this mesa, is obtained by counting
branches. A rough estimation, valid for all our mesas
is $N\simeq V_{max}/10$ mV, where $V_{max}$ is the voltage at a
maximum in the $I$-$V$.
The maximum is caused by back-bending
at high bias due to self-heating. The extent of self-heating
depends on geometry and decreases with decreasing mesa sizes
\cite{Heating_2001,Heating_2005}. Sizes of our mesas are in the
range $(2.5-10)\times (10-30)~\mu$m$^2$ with areas
$50<A<300~\mu$m$^2$, which are significantly smaller than ``large"
mesas ($A>10^4~\mu$m$^2$) studied in the majority of earlier works
\cite{Ozyuzer_2007,Benseman_2013,Welp_2013,Kashiwagi_2015,HBWang_2015,Sun_2018,Kashiwagi_2018,HBWang_2019,Kuwano_2020,Tsujimoto_2020,Saiwai_2020,Delfanazari_2020},
but larger than ``small" mesas ($A\lesssim 10~\mu$m$^2$) studied
in Ref. \cite{Borodianskyi_2017}. Thus, our mesas are of
``intermediate size". This enables a significant overall power
$\sim 1$ mW with a tolerable self-heating.

\begin{figure*}[th]
    \centering
    \includegraphics[width=0.99\textwidth]{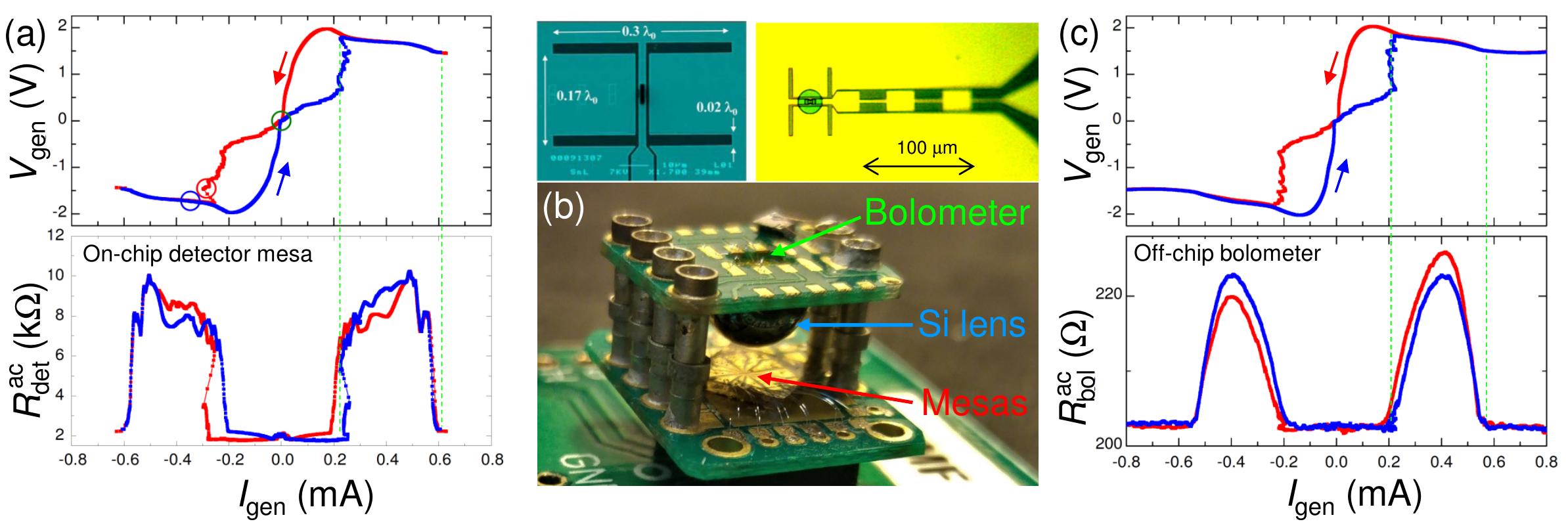}
    \caption{(Color online). Comparison of on-chip and off-chip detection schemes.
    (a) On-chip generation-detection with electrically connected
    mesas. Top panel: the $V$-$I$ of a generator mesa at $T\simeq 3$ K. Circles represent bias points for the detector $I$-$V$'s in Fig. \ref{fig:fig1} (d).
    Bottom: ac-resistance of the detector mesa as a function of the
    generator current. Blue/red curves represent up and down bias sweeps.
    (b) A setup for off-chip detection of emission by a NbN bolometer and images of the bolometer (from Refs. \cite{Cherednichenko_2008,Motzkau_PhD}).
    (c) Off-chip detection experiment. Top: The $V$-$I$ of the generator mesa. Bottom: the bolometer response as a function
    of $I_{gen}$. Vertical dashed lines in (a) and (c) indicate that emission occurs at the step in the generator $I$-$V$.
    }
    \label{fig:fig2}
\end{figure*}

To analyze EMW emission, we start with an on-chip
generation-detection scheme \cite{Borodianskyi_2017} using one
mesa as a generator and another mesa on the same chip as a
switching-current detector. Figure \ref{fig:fig1} (d) shows
evolution of the $I$-$V$'s of the detector at three bias points
$(I_{gen}, V_{gen})$ in the generator. The $I$-$V$ of the
generator mesa, with marked bias points, is shown in the top panel
of Fig. \ref{fig:fig2} (a). From Fig. \ref{fig:fig1} (d) it is
seen that with increasing bias in the generator, critical currents
in the detector are first reduced (red curve) and then get
completely suppressed (blue curve) due to the EMW absorption
\cite{Borodianskyi_2017}. Detailed dynamics of the
generation-detection experiment is demonstrated in the
supplementary video \cite{Supplem}.

Absorption of EMW by a JJ leads to formation of Shapiro steps in
the $I$-$V$ at $V_n=nhf/2e$ ($n$-integer). Thus, junction response
carries a spectroscopic information both about the frequency and
the amplitude of EMW. 
In mesas with $N$ JJ's, the EMW emission occurs at
$f_J=2eV_{gen}/Nh$ (provided all JJ's are synchronized). Thus, the
primary Shapiro step should appear at $V_{det}=V_{gen}/N$. Small
steps can indeed be seen in the blue $I$-$V$ from Fig.
\ref{fig:fig1} (d). Fig. \ref{fig:fig1} (e) shows the differential
conductance for this curve. It exhibits clear peaks, corresponding
to steps in the $I$-$V$. The grid spacing is equal to $V_{gen}/N$
with estimated $N\simeq 200$ for the generator mesa, see Fig.
\ref{fig:fig2}(a), and is in agreement with the observed peak
separation. Some displacement of peak positions is likely caused
by the fact that the detector response involves six JJ's, see Fig.
\ref{fig:fig1} (d). A certain difference between JJ's leads to a
mismatch of bias conditions for appearance of Shapiro steps. This
affects both the regularity and the amplitude of Shapiro steps.
Nevertheless, the data is consistent with occurrence of
monochromatic emission at $f_J\simeq 4.3$ THz because the
non-monochromatic emission would not lead to appearance of
distinct steps.

To exclude possible electrical crosstalk, we performed similar
generation-detection experiment using electrically disconnected
mesas. For this the whisker was cut by a focused ion beam in two
sections, thus separating the generator and the detector mesas.
The detector response remains qualitatively unchanged after such
separation. Fig. \ref{fig:fig1} (f) shows the $I$-$V$ of the
generator mesa, color maped by the response of electrically
disconnected detector mesa (ac-resistance, $R_{det}^{ac}$,
measured by the lock-in technique with zero offset).
It is seen that 
a profound, almost vertical step develops in
the back-bending region of the $I$-$V$. 
The color code indicates that the detector response appears just at this step and disappears both above and
below. In Fig. \ref{fig:fig2} (a) the same behavior is
demonstrated for the on-chip generation-detection experiment with
electrically connected mesas. Here a correlation between the step
in the $I$-$V$ of the generator and the upturn in the
detector response is clearly seen (see also the Supplementary
video \cite{Supplem}). Such a non-monotonous behavior precludes
self-heating origin of the observed signal and confirms occurrence
of the EMW emission \cite{Borodianskyi_2017}.

To confirm EMW emission into open space we perform off-chip
detection using a NbN bolometer
\cite{Cherednichenko_2008}. Fig. \ref{fig:fig2} (b) shows the
measurement setup and images of the bolometer (from Refs.
\cite{Cherednichenko_2008,Motzkau_PhD}).
The bolometer is placed at a distance $\sim 1$ cm above the
device. Fig. \ref{fig:fig2} (c) shows corresponding generation
(top) and detection (bottom) characteristics. Vertical dashed
lines indicate that the bolometer response (lock-in resistance
$R_{bol}^{ac}$) appears at the step in $I_{gen}$-$V_{gen}$, thus
confirming EMW emission. Some difference in the shapes of on-chip,
Fig. \ref{fig:fig2} (a), and off-chip, Fig. \ref{fig:fig2} (c),
responses is likely caused by the spectral selectivity of the
double-slot antenna of the bolometer, peaked at $f\simeq 1.6$ THz,
well below the emission frequency $f_J\simeq 4.2$ THz for this
mesa \cite{Note3}.

\begin{figure*}[t]
    \centering
    \includegraphics[width=0.99\textwidth]{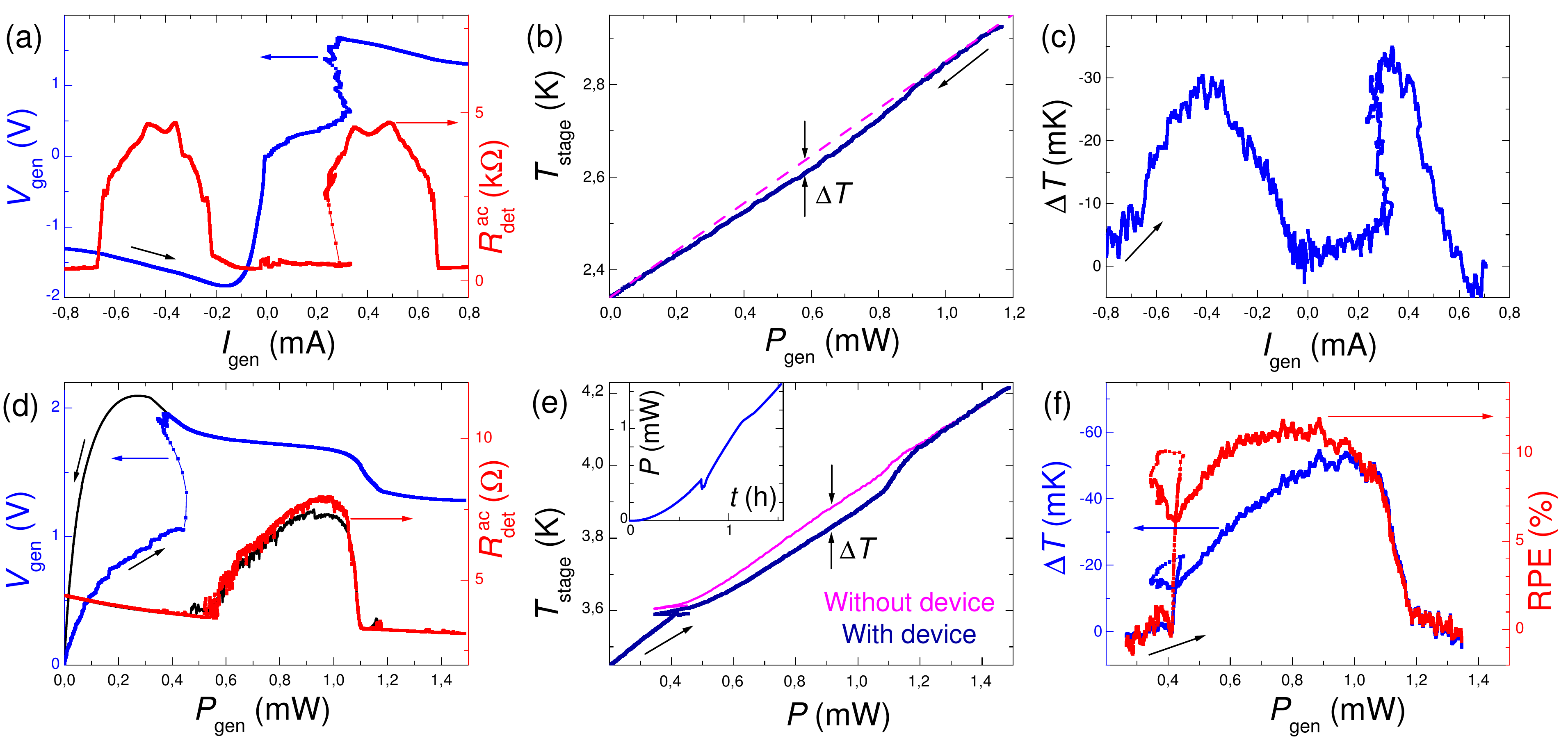}
    \caption{ (Color online). Radiative cooling analysis of emission for two devices (a-c) and (d-f). (a) The generator $V$-$I$ (blue, left axis) and the on-chip detector response (red, right axis) vs. $I_{gen}$. (b) The stage temperature (dark blue) vs. the dissipation power $P_{gen}$ for an upward bias sweep ($I_{gen}<0$). It deviates from the linear dependence (magenta dashed
line) at $P_{gen}\simeq 0.6$ mW. (c) The temperature drop vs. $I_{gen}$. It is seen that $\Delta T$ is correlated with the detector
response in (a). (d) The generator mesa voltage (blue, left axis) and
the detector response (red, right axis) vs. $P_{gen}$ for an upward bias sweep at another device. Black lines correspond to the reverse sweep. (e) $T_{stage}$ vs. the dissipation power for experiments with (dark blue) and without (magenta) device for upward bias sweeps. The inset shows time dependencies of the dissipation power, identical for both cases. (f) Temperature drop (blue, left axis) and radiation power efficiency (red, right axis) vs $P_{gen}$. It is seen that RPE reaches $12\%$.
    }
    \label{fig:fig3}
\end{figure*}

Power dissipation leads to heating, but EMW emission -- to
radiative cooling of the device. We use this for unambiguous
estimation of the emission power, $P_{THz}$. Figure \ref{fig:fig3}
(a) represents the on-chip generation-detection measurement for
another device (upward bias sweep). The sample stage of our dry
cryostate has a relatively small heat capacitance and a finite
heat resistance, $R_{th}$, to the coldhead. Therefore, its
temperature directly reflects the energy balance at the chip. Fig.
\ref{fig:fig3} (b) shows the stage temperature as a function of
the dissipation power, $P_{gen}=I_{gen} V_{gen}$, for the data
from Fig. \ref{fig:fig3} (a). As expected from the Newton's law of
cooling, it increases approximately linearly,
$T_{stage} \simeq T_0+R_{th} P_{gen}$, with $R_{th}=0.508$ K/mW.
However, at $P_{gen}\simeq 0.6$ mW there is a visible drop down
from the linear dependence. 
In Fig. \ref{fig:fig3} (c) we plot this drop, $\Delta T$, vs.
$I_{gen}$. It clearly follows the detector response, shown by the
red line in Fig. \ref{fig:fig3} (a). The correlation between
$\Delta T$ and the EMW emission provides a straightforward
demonstration of the radiative cooling phenomenon. Importantly, it
allows direct estimation of $P_{THz}= \Delta T/R_{th}$. The
maximum drop $\Delta T = -31.5 \pm 2$ mK occurs at $I_{gen}\simeq
0.4$ mA and $P_{gen}=0.6$ mW. This yields $P_{THz} \simeq 62~\mu$W
and RPE of $10.3\pm 0.7 \%$ for this device.

Figs. \ref{fig:fig3} (d-f) show similar measurements for another
device. Fig. \ref{fig:fig3} (d) shows $V_{gen}$ (blue, left axis)
and $R_{det}^{ac}$ (red, right axis) vs. $P_{gen}$ for the upward
bias sweep (black lines represent similar data for the reverse
sweep). The dark blue curve in Fig. \ref{fig:fig3} (e) shows
$T_{stage}$ vs. $P_{gen}$ for the upward sweep. Inset shows time
dependence of $P_{gen}(t)$ during this sweep. It is nonlinear due
to the non-Ohmic $I$-$V$ of the generator. The nonlinearity may
cause some transient effects.
Therefore, to make an accurate calibration of the bare (without
device) thermal response of the stage we heated it by a resistor,
applying exactly the same time-dependence of the
dissipation power 
using a programmable current source. The corresponding variation
of $T_{stage}$ is shown by the magenta line in Fig. \ref{fig:fig3}
(e). It is almost linear with the slope $R_{th}\simeq 0.514$ K/mW.
Fig. \ref{fig:fig3} (f) shows the difference between $T_{stage}$
with and without device (blue) and the radiation power efficiency
RPE$=-\Delta T/(R_{th} P_{gen})$ (red curve). It is seen that the
maximum RPE for this device reaches $12\%$, which corresponds to
the emission power of $P_{THz}\simeq 0.11$ mW.
We want to emphasize that this power is actually emitted in the
far-field. Indeed, it is taken out of the range cooled by the
second stage of the cryocooler (which has a small cooling power)
and is dumped into the first stage (with a very large cooling
power), which is 10-40 cm away. 
This distance is much larger that the wavelength, $\lambda\sim 70~
\mu$m, at $f\sim 4.3$ THz. Therefore, the radiative cooling probes
the total far-field emission. Of course, for the practical device
the emission should be collected and delivered via open space to
the desired place, which inevitably involve certain losses. In
this respect the estimation of RPE from the radiative cooling
provides the upper limit of the achievable ``useful" emission.

Estimations above indicate that whisker-based THz sources can have
RPE $>10\%$, not far from the theoretical limit of $50\%$
\cite{Krasnov_2010}. This significantly exceeds the RPE for
devices made on regular Bi-2212 crystals
\cite{Benseman_2013,Borodianskyi_2017}.
Generally, the low RPE is caused by poor impedance matching with open space 
\cite{Krasnov_2010}. JJ's are much smaller than the wave length.
Especially tiny are interlayer distances $\sim 1$ nm. Therefore,
JJ's act as miniature dipoles with almost no emission in the far
field. Essentially, the emission is facilitated by other passive
but large-size elements of the device such as the Bi-2212 crystal
and the electrodes, acting as matching antennas. Optimization of
emission requires proper microwave design for impedance matching.
In the Supplementary \cite{Supplem} we discuss the geometrical
differences between crystal- and whisker-based devices. We argue
that the reported boost of RPE in whisker-based devices is cause
by a specific turnstile antenna-like geometry, which allows
obviation of a large parasitic capacitance between the crystal and
the electrode and facilitates good impedance matching with open
space.


To conclude, a low radiation power efficiency is one of the main
problems of THz sources. For cryogenic sources the emission power
is limited both by the cooling power and RPE: $P_{THz} <RPE \times
P_{cooling}$. For portable cryocoolers with limited $P_{cooling}$,
the only way to increase the emission power is by enhancement of
the RPE.
We presented prototypes of novel THz sources based on
Bi$_2$Sr$_2$CaCu$_2$O$_{8+\delta}$ whiskers
with RPE up to $12\%$, more than an order of magnitude larger than
for similar devices made on regular Bi-2212 crystals. This
indicates better impedance matching with open space due to the
specific turnstile antenna-like geometry of whisker-based devices.
We argue that such devices can be used for creation of tunable,
monochromatic, continuous-wave, compact and, most importantly,
power-efficient THz sources.

\begin{acknowledgments}
The work was supported by the Russian Science Foundation Grant No.
19-19-00594. We are grateful to A. Agostino and M. Truccato for
assistance with whisker preparation, to S. Cherednichenko for
providing the bolometer and to A. Efimov and K. Shiianov for
assistance in experiment. The manuscript was written during a
sabbatical semester of V.M.K. at MIPT.
\end{acknowledgments}


%

\end{document}